\documentclass[hyper]{JHEP3}

\keywords{Field Theories in Lower Dimension, Soliton Monopoles and Instantons, Matrix Models}

\skip\footins = 1\bigskipamount plus 2pt minus 4pt

\usepackage{epsfig}
\newcommand{\be}{\begin{equation}}
\newcommand{\ee}{\end{equation}}
\newcommand{\bea}{\begin{eqnarray}}
\newcommand{\eea}{\end{eqnarray}}

\title{Comment on "Solitons and excitations in the
duality-based matrix model"}

\author{V. Bardek$^{\dagger}$ and S. Meljanac$^{\dagger}$\\ 

$^{\dagger}$ 
Institute Rudjer Bo\v skovi\'c, Bijeni\v cka cesta 54,\\
P.O. Box 180, HR-10002 Zagreb, Croatia\\
E-mail: \email{bardek@irb.hr},
\email{meljanac@irb.hr}}

\abstract{
It is shown that a method for constructing
exact multi-solitonic solutions of the coupled BPS
equations in the duality-based generalization of the
hermitean matrix model, which was put forward in a
recent paper, is not correct.}

\begin{document}


In the paper JHEP {\bf 08}, 064 (2005) \cite{AJJ}, 
Andri\'c et al. claim that a specific, duality-based
generalization of the hermitean matrix model possesses
topologically non-trivial, multi-solitonic solutions.
In the BPS limit, they have constructed two-soliton and
n-soliton solutions to the coupled non-linear equations.
Unfortunately, the purpose of this Comment is to point
out an error in that derivation.

Following Ref. \cite{AJJ}, we write the coupled BPS equations
(Eqs. (2.4) and (2.5) in Ref. \cite{AJJ}) as
\begin{eqnarray}
\frac{\lambda-1}{2}\frac{\partial_x\rho(x)}{\rho(x)}
+ \lambda\int{\mbox{\hspace{-1.05em}{\footnotesize$\diagup$}}}\: \frac{dy\, \rho(y)}{x-y}
+ \int{\mbox{\hspace{-1.05em}{\footnotesize$\diagup$}}}\: \frac{dy\, m(y)}{x-y}
&=& 0\: ,
\label{BPS1}
\\
\frac{1-\lambda}{2}\frac{\partial_x m(x)}{m(x)}
+ \lambda\int{\mbox{\hspace{-1.05em}{\footnotesize$\diagup$}}}\: \frac{dy\, \rho(y)}{x-y}
+ \int{\mbox{\hspace{-1.05em}{\footnotesize$\diagup$}}}\: \frac{dy\, m(y)}{x-y}
&=& 0\: .
\label{BPS2}
\end{eqnarray}
The solutions to these equations are always interrelated by 
\begin{equation}
\rho(x) m(x) = c,
\label{mc_rel}
\end{equation}
where $c$ is some positive constant. This means that there is only one relevant 
equation since the other one is generated by the substitution $\rho(x) = \frac{c}{m(x)}$.
The key point missed by Andri\'c et al. is the fact that the equivalence of 
(\ref{BPS1}) and (\ref{BPS2}) was established on the basis of interchanging 
$\rho$ by $\frac{c}{m}$ (or vice versa), but only for $c\ne 0$.

Generally, for $c=0$, the two equations are completely independent of each other,
and must be simultaneously considered in order to get the right solutions.
For the sake of completeness, we now briefly recall the main findings concerning
the new soliton solutions of Ref. \cite{AJJ}, and clarify some points which were not
made explicitly clear in Ref. \cite{AJJ}. Let us start with 
the one-soliton solutions of the coupled
equations (\ref{BPS1}) and (\ref{BPS2}). The solutions (Eqs. (2.12) and (2.13) of
Ref. \cite{AJJ}) are given by
\begin{eqnarray}
\rho(x) &=& \rho_0 \frac{x^2 + a^2}{x^2 + b^2},
\nonumber\\
m(x) &=& \frac{c}{\rho_0} \frac{x^2 + b^2}{x^2 + a^2}. 
\label{BPSsolAJJ}
\end{eqnarray}
These equations have already been 
found and discussed in our earlier work on the two-family 
Calogero systems (see Ref. \cite{BM1} and published paper \cite{BM2}).
Here, we write down the values of the parameters $a$, $b$ and $c$
(they were not calculated in Ref. \cite{AJJ}):
\begin{eqnarray}
a &=& 
\frac{N (\lambda -1)}{\rho_0 \pi M \left(1 - \frac{N^2}{M^2} \lambda^2 \right)}\: ,
\nonumber\\
b &=&
\frac{1 - \lambda}{\lambda \rho_0 \pi 
\left(1 - \frac{M^2}{N^2} \frac{1}{\lambda^2} \right)}\: ,
\nonumber\\
c &=& \rho_0^2 \frac{M}{N}.
\label{BPS_BM}
\end{eqnarray}
At this stage it is quite obvious that, if the number of particles in the first 
family $N$ is much larger than the number of particles in the second family $M$,
the interesting limiting case, $c\to 0$ (which implies simultaneously $a\to 0$
by virtue of Eq. (\ref{BPS_BM})), appears. In this limit, 
which is also considered in our work \cite{BM1} and paper \cite{BM2}, the 
first-family soliton reduces to the vortex profile,
\begin{equation}
\rho(x)\ =\ \rho_0 \frac{x^2}{x^2+\left(\frac{1-\lambda}{\lambda \pi \rho_0}\right)^2}\: ,
\end{equation}
while the second-family soliton transforms into the delta-function,
\begin{equation}
m(x)\ =\ (1 - \lambda) \delta(x).
\end{equation}
Let us stress that the product of $\rho$ and $m$ vanishes.
The authors of Ref. \cite{AJJ} checked that $m(x)$ indeed satisfies 
the second equation (\ref{BPS2}). After all, this is not surprising since every 
member of the infinite set of approaching solutions $m(x)$ (corresponding 
to infinitesimal values of the parameter $c$) satisfies 
the same equation from the very beginning.

We point out that vortices of this type can never occur in the
generalized hermitean matrix model of Ref. \cite{AJJ}. Namely,
this model can be interpreted as a two-family Calogero model for
a special choice of the coupling strength $\lambda = \frac{1}{2}$ and
$\lambda = 2$
 (for dual family), corresponding to symmetric and real quaternionic
matrices, respectively. In this case, the ratio of ranks of matrices
$\frac{M}{N}$ is equal $\frac{1}{2}$, and consequently cannot be made arbitrarily
small (see Eq. (\ref{BPS_BM})). Moreover, in this case $\rho(x) = \rho_0$.

Let us now turn our attention to the completely different situation concerning 
the so-called two-soliton solutions (see Eq. (2.21) of Ref. \cite{AJJ}).  
The Ansatz proposed in Ref. \cite{AJJ} is exactly the same as ours, already given in 
Ref. \cite{BM1} (see Eq. (34)), i.e. it is the rational function of two
polynomials with the fourth degree in $x$, in principle ensuring 
the two-hole (or for $m=\frac{c}{\rho}$, the two-lump) structure. This 
Ansatz,
\begin{equation}
\rho(x)\ =\ \rho_0 \frac{(x^2-a^2)(x^2-\overline{a}^2)}
                        {(x^2-b^2)(x^2-\overline{b}^2)}\: ,
\qquad m(x)=\frac{c}{\rho(x)}\: ,
\label{TS_ans}
\end{equation}
takes into account the positiveness of the particle density automatically.
It is somehow a natural generalization of the one-soliton solution 
(\ref{BPSsolAJJ}). Unfortunately, there are no nontrivial solutions for 
the complex parameters $a$, $\overline{a}$, $b$, $\overline{b}$, and for 
the real parameter $c$ (hence nothing analogous to Eq. (\ref{BPS_BM})).
For more details, see Ref. \cite{BM1}, where we discuss the algebraic implications 
of the Ansatz (\ref{TS_ans}). The fact that there are no solutions 
of the form (\ref{TS_ans}) for $c\ne 0$ was also recognized by the authors 
of Ref. \cite{AJJ}. Nevertheless, they suggested that the expression 
(\ref{TS_ans}) could be viewed as a genuine solution of the system 
(\ref{BPS1}) and (\ref{BPS2}), but only in the $c\to 0$ limit!
After performing the simultaneous limiting procedure $c\to 0$ and ${\cal I}m\, a \to 0$ 
in the expressions for $\rho$ and $m$ (Eq. (\ref{TS_ans})), the authors 
of Ref. $\cite{AJJ}$ claimed that 
\begin{equation}
\rho(x)\ =\ \rho_0 \frac{(x^2-x_0^2)^2}
                        {(x^2-b^2)(x^2-\overline{b}^2)}
\label{rhoAJJ}
\end{equation}
and 
\begin{eqnarray}
m(x) &=& (1-\lambda)(\delta(x-x_0) + \delta(x+x_0))\: ,
\label{mAJJ}
\\
x_0 &=& {\cal R}e\, a\; ,
\end{eqnarray}
were the topologically non-trivial, solitonic solutions of 
equations (\ref{BPS1}) and (\ref{BPS2}).
This time, the check of the validity of the second equation (\ref{BPS2})
was completely ingnored in Ref. \cite{AJJ}! However, this has to be done for the sake of
consistency.
Inserting these expressions into the system (\ref{BPS1}) and (\ref{BPS2}),
we find that equation (\ref{BPS1}) is identically satisfied, whereas 
equation (\ref{BPS2}) is satisfied if
\begin{equation}
[\delta(x-x_0) - \delta(x+x_0)]
\left(
\frac{1-\lambda}{2 x_0}
+ \lambda
\int{\mbox{\hspace{-1.05em}{\footnotesize$\diagup$}}}\: 
\frac{dy \rho(y)}{x_0 - y} 
\right)\: \ =\ 0\: ,
\end{equation}
which implies that 
\begin{equation}
\frac{1}{x_0^2 - b^2} + \frac{1}{x_0^2 - \overline{b}^2}\ =\ \frac{1}{2 x_0^2}.
\label{cond1}
\end{equation}
Of course, this is to be expected! Equations (\ref{BPS1}) and (\ref{BPS2})
are completely independent for $c=0$. In this case, contrary to what we have in the 
one-soliton sector, there are no approaching solutions (i.e. solutions for 
infinitesimal $c$ and ${{\cal I}m\: a})$ which would render equation 
(\ref{BPS2}) superfluous.
After introducing 
\begin{equation}
b\ =\ |b| e^{i\varphi}\qquad \mbox{and} \qquad r\ =\ \frac{x_0^2}{|b|^2}
\end{equation}
into the condition (\ref{cond1}), we obtain
\begin{equation}
3 r^2 - 2 r \cos\, 2\varphi - 1 \ =\ 0. 
\end{equation}
Finally, this condition and the similar one (2.26) from Ref. \cite{AJJ},
\begin{equation}
r - 2 r \cos\,\varphi + 1 \ =\ 0,
\end{equation}
lead to 
\begin{equation}
r^2\ =\ 1 \qquad \mbox{and} \qquad \cos\,\varphi\ =\ 1,
\end{equation}
which means that $x_0^2 = b^2$, and this brings us back to the constant 
solution $\rho = \rho_0$. However, because of the constraint
\begin{equation}
\rho(x) m(x)\ =\ 0,
\end{equation}
this is possible only for $\rho_0 = 0$ or $\lambda = 1$,
which is of course unacceptable. 
Therefore, the proposed functions $\rho$ and $m$ from Eqs. (\ref{rhoAJJ}) and
(\ref{mAJJ}) are not the solutions of the BPS equations (\ref{BPS1}) and
(\ref{BPS2}). Consequently, nor are the $n$-soliton generalizations
given by Eq. (2.29) in Ref. \cite{AJJ}.

In conclusion, the approach suggested in Ref. \cite{AJJ} is not correct.
The problem of obtaining closed-form multi-soliton solutions of 
the coupled BPS equations (\ref{BPS1}) and (\ref{BPS2}) remains 
unsolved.\\[1cm]
\acknowledgments

This work is supported by the Ministry of Science and Technology of 
the Republic of Croatia under contract No. 0098003.

\end{document}